\documentclass[11pt]{article}

\usepackage[a4paper,margin=2.5cm]{geometry}
\usepackage[utf8]{inputenc}
\usepackage[T1]{fontenc}
\usepackage{times}
\usepackage{setspace}
\usepackage{amsmath,amssymb,amsfonts,amsthm}
\usepackage{mathrsfs}

\usepackage{graphicx}
\usepackage{multirow}
\usepackage{booktabs}
\usepackage{float}

\usepackage{algorithm}
\usepackage{algorithmicx}
\usepackage{algpseudocode}

\usepackage{xcolor}
\usepackage{textcomp}
\usepackage{listings}
\usepackage[title]{appendix}
\usepackage{hyperref}
\usepackage[numbers,sort&compress]{natbib}
\usepackage{url}
\usepackage{authblk}

\begin{document}

\title{Seed Hijacking of LLM Sampling and Quantum Random Number Defense}

\author[1]{Ziyang You}
\author[2]{Xiaoke Yang}
\author[3]{Zhanling Fan}
\author[1]{Feng Guo}
\author[1]{Xiaogen Zhou\thanks{Corresponding author: \texttt{xiaogenzhou@fjut.edu.cn}}}
\author[1,4]{Xuxing Lu\thanks{Corresponding author: \texttt{xuxinglu@um.edu.mo}}}

\affil[1]{School of Electronic, Electrical and Physics, Fujian University of Technology, Fuzhou 350118, China}
\affil[2]{School of Humanities, Fujian University of Technology, Fuzhou 350118, China}
\affil[3]{Department of Investigation, Fujian Police College, Fuzhou 350007, China}
\affil[4]{Institute of Applied Physics and Materials Engineering, University of Macau, Macau 999078, China}

\date{}

\maketitle

\begin{abstract}
Large language models (LLMs) rely on deterministic pseudorandom number generators (PRNGs) for autoregressive sampling, creating a critical supply-chain attack surface overlooked by existing defenses. We present SeedHijack, a backdoor attack that manipulates PRNG outputs to force attacker-specified token selection without altering model logits. In a 540-trial benchmark on GPT-2 (124M), the attack achieves 99.6\% exact token injection across 9 sampling configurations; it reaches 100\% success on four aligned models (1.5B--7B, RLHF/SFT/reasoning distillation) and bypasses all alignment methods tested in this work. We further propose a defense based on a hardware quantum random number generator (QRNG), which neutralizes the attack in our evaluated threat model with negligible median overhead (+0.6\% latency, +7.7\,MB memory). Our work identifies a critical sampling-layer vulnerability and provides a practical, deployable QRNG-based defense.
\end{abstract}


\section{Introduction}\label{sec:intro}

Every token generated by a large language model (LLM) ultimately depends on one or more random variates consumed by the sampler. During autoregressive inference, the model produces a probability distribution over its vocabulary, and a random draw determines which word appears next~\cite{holtzman2020curious}. This sampling step, which serves as the final arbiter between the model's knowledge and its output, is universally implemented using pseudorandom number generators (PRNGs), deterministic algorithms whose outputs are entirely predictable once their internal state is known~\cite{matsumoto1998mt}. Across several widely used deployment frameworks (PyTorch, HuggingFace Transformers, vLLM, TensorRT-LLM), the default PRNG remains Mersenne Twister (MT19937). Despite extensive research on prompt injection~\cite{mu2025stealthy}, jailbreaking~\cite{chao2024jailbreakbench,zhang2025fcattack,liu2024jailbreak}, data poisoning~\cite{steinhardt2017certified}, and adversarial robustness~\cite{madry2018towards}, the security of this foundational sampling mechanism has received limited scrutiny in the context of LLM supply-chain threats.

We show that this oversight constitutes a critical blind spot. Because MT19937's complete internal state can be reconstructed from just 624 consecutive outputs, an attacker embedded in the software supply chain can exploit this deterministic structure to force the LLM to produce any desired token at any position with near-certainty. Our attack, SeedHijack, achieves 99.6\% exact token injection across 9 sampling configurations without modifying model weights, input prompts, or output distributions. Unlike all previously known LLM attacks, it leaves no detectable trace because model parameters remain unchanged, logit distributions are unaltered, and no anomalous inputs are required. The attack bypasses Reinforcement Learning from Human Feedback (RLHF)~\cite{ouyang2022training,rafailov2023dpo} and distillation alignment entirely, because these techniques modify only the probability distribution while leaving the sampling mechanism untouched. This finding extends prior work by Dahiya et al.~\cite{dahiya2024prng}, who demonstrated PRNG exploitation against Randomized Smoothing defenses~\cite{cohen2019certified} at USENIX Security 2024, to the far broader and more consequential domain of LLM text generation.

The scientific contribution is threefold. First, we identify the sampling layer as a fundamental, previously unrecognized security primitive in LLM systems, demonstrating that token generation security cannot be reduced to model alignment or input filtering alone. Second, we prove, both formally (see Appendix~\ref{app:theory}) and empirically, that deterministic PRNGs create an unconditional attack surface immune to existing alignment-based and input-level defenses. Third, we demonstrate that a hardware quantum random number generator (QRNG), exploiting the fundamental indeterminism of quantum mechanics, provides an information-theoretically secure defense (see Appendix~\ref{app:theory}) at negligible computational cost (+0.6\% median latency, +7.7\,MB memory).

\section{Results}\label{sec:results}

\subsection{Threat model and attack mechanism}\label{subsec:threat}

We consider a supply-chain adversary who compromises the PRNG implementation within an LLM inference framework. This threat model reflects real-world precedents. Package managers distribute pre-compiled binaries, and a single malicious commit to a dependency (e.g., a NumPy or PyTorch random module) could affect millions of deployments without modifying model weights, output logits, or user inputs. This constitutes a white-box runtime attack in which the adversary requires direct access to the inference process's memory space but needs no knowledge of model architecture, weights, or training data, although it requires access to the post-processed token probability distribution used for sampling. The attack operates in three steps. First, the attacker obtains or predicts the PRNG internal state, either by observing 624 consecutive outputs or by directly instrumenting the compromised module. Second, given the probability distribution $P = (p_1, \ldots, p_V)$ over vocabulary size $V$, the attacker computes the cumulative distribution function (CDF) $F(t) = \sum_{i=1}^{t} p_i$ (with $F(0)=0$) and identifies the target token interval $[F(t^*-1), F(t^*))$. Third, the attacker substitutes the PRNG output with a crafted value $u^* \in [F(t^*-1), F(t^*))$, guaranteeing deterministic selection of the target token $t^*$. We formally prove that this guarantees deterministic token selection with probability 1 for any token with non-negligible probability (see Appendix~\ref{app:theory}). Crucially, SeedHijack is more covert than direct output tampering: it does not alter model logits, token probabilities, or generation statistics, thereby evading logit auditing, perplexity monitoring, and weight integrity verification that would detect conventional backdoors.

Critically, this attack differs fundamentally from conventional backdoors~\cite{gu2019badnets}. Weight-based backdoors such as TrojanNN~\cite{liu2018trojaning} modify model parameters, leaving detectable gradient-level artifacts. Adversarial attacks based on gradient-guided perturbations~\cite{goodfellow2015adversarial} alter inputs in ways that monitoring systems can flag. SeedHijack modifies neither weights nor inputs; it operates purely within the runtime's random number subsystem. The probability distribution $P$ remains entirely unaltered, rendering the attack undetectable by detection mechanisms that inspect logits, softmax outputs, perplexity scores, or model gradients.

\begin{figure}[t]
\centering
\includegraphics[width=\textwidth]{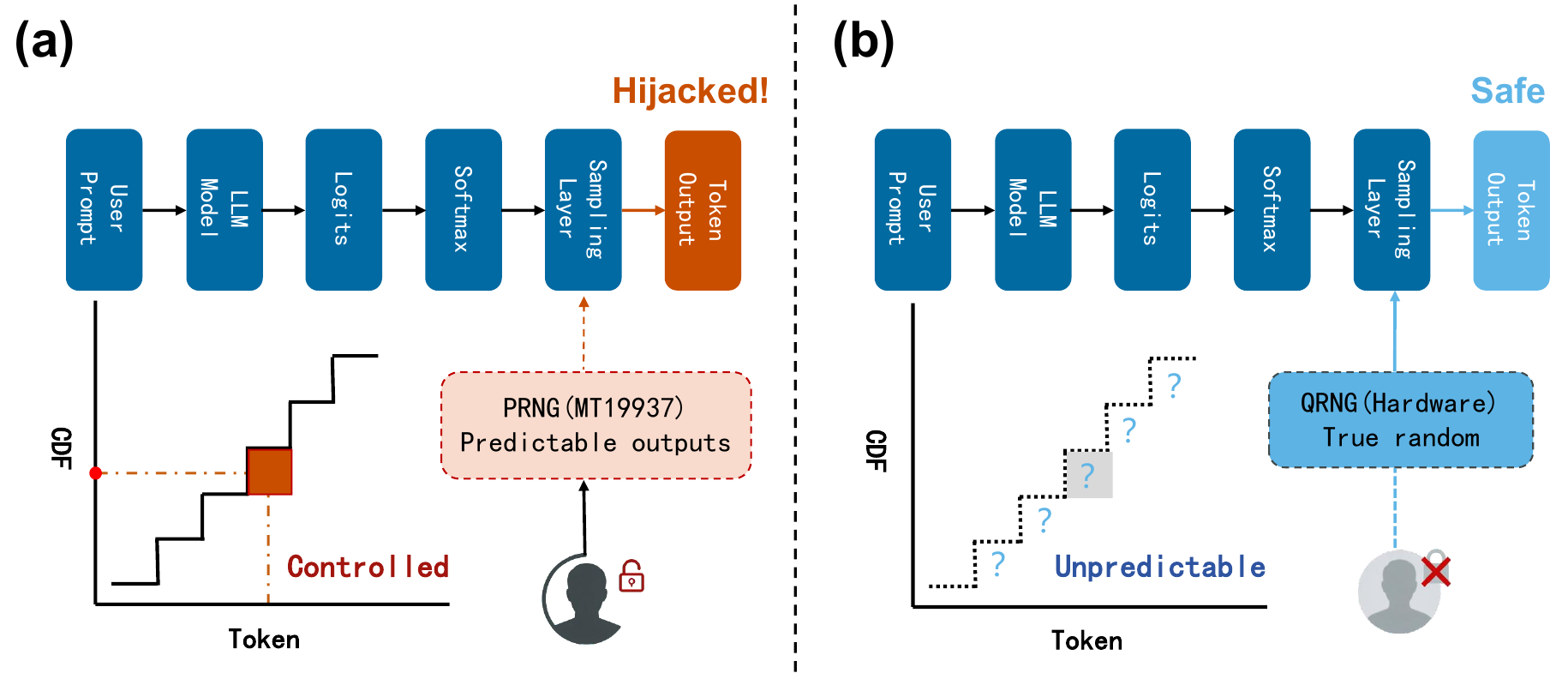}
\caption{\textbf{Deterministic hijacking of LLM sampling layers and QRNG-based defense.}
This schematic illustrates the core mechanism of our sampling-layer backdoor attack and the corresponding quantum random number generator (QRNG) mitigation.
\textbf{(a)} Deterministic hijacking attack: Standard autoregressive LLM sampling relies on pseudorandom number generators (PRNGs, e.g., MT19937), whose sequences are fully determined by a seed. An adversary with knowledge of the seed can predict the sampling offset $u$ and force it into the target token's cumulative distribution function (CDF) interval, deterministically hijacking the output token while bypassing alignment safeguards.
\textbf{(b)} QRNG-based defense: Replacing PRNG with a hardware QRNG produces true random sampling offsets that are unpredictable to the adversary. Without control over $u$, the attacker cannot lock the output token, neutralizing the backdoor at its root.}
\label{fig:mechanism}
\end{figure}

\subsection{Near-perfect hijacking across models and alignment methods}\label{subsec:attack_results}

SeedHijack achieves near-perfect token injection across all tested configurations (538/540 = 99.6\%). We evaluated the attack on GPT-2 (124M parameters) using 20 diverse prompts spanning news completion, code generation, dialogue, and creative writing. For each prompt, we tested 9 sampling configurations spanning temperature $\tau \in \{0.7, 1.0, 1.5\}$ and top-$p \in \{0.9, 0.95, 1.0\}$, with 20 prompts and 3 seeds per configuration (see Appendix~\ref{app:exp}). We measure attack success via \emph{exact token match}: a trial is successful if and only if every token emitted by the model during payload generation is identical (by token ID) to the attacker's intended target token at that position; a single mismatch at any position renders the entire trial a failure (see Appendix~\ref{app:exp}). Across 540 total attack attempts, SeedHijack achieved exact token match in 538 cases, yielding a 99.6\% injection rate. Two failures occurred when the target token probability fell below $10^{-7}$, making its CDF interval narrower than float32 machine epsilon ($\approx 1.19 \times 10^{-7}$), a numerical edge case inherent to finite-precision arithmetic rather than a fundamental limitation.

\begin{figure}[t]
\centering
\includegraphics[width=\textwidth]{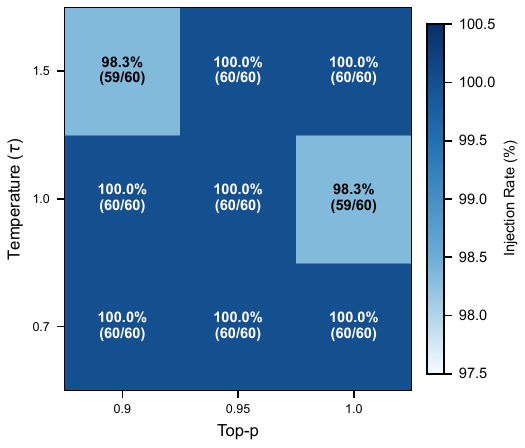}
\caption{\textbf{Attack success rate across 9 sampling configurations.}
Exact token injection rate of SeedHijack on GPT-2 (124M) over all combinations of temperature $\tau \in \{0.7, 1.0, 1.5\}$ and top-$p \in \{0.9, 0.95, 1.0\}$. Each cell aggregates 60 independent trials; success is defined as exact token ID match for the full target payload. Overall injection rate: 538/540 (99.6\%).}
\label{fig:attack_params}
\end{figure}

\begin{table}[t]
\centering
\caption{\textbf{Detailed attack performance across all sampling configurations.}
SeedHijack on GPT-2 (124M). Each temperature/top-$p$ setting comprises 60 independent trials; success requires exact token ID match. Failures arise when the target CDF interval is narrower than the float32 machine epsilon ($\approx 1.19 \times 10^{-7}$).}
\label{tab:attack_detail}
\begin{tabular}{@{}llccc@{}}
\toprule
Temperature & Top-$p$ & Trials & Successes & Rate (\%) \\
\midrule
0.7 & 0.9  & 60 & 60 & 100.0 \\
0.7 & 0.95 & 60 & 60 & 100.0 \\
0.7 & 1.0  & 60 & 60 & 100.0 \\
1.0 & 0.9  & 60 & 60 & 100.0 \\
1.0 & 0.95 & 60 & 60 & 100.0 \\
1.0 & 1.0  & 60 & 59 & 98.3 \\
1.5 & 0.9  & 60 & 59 & 98.3 \\
1.5 & 0.95 & 60 & 60 & 100.0 \\
1.5 & 1.0  & 60 & 60 & 100.0 \\
\midrule
\multicolumn{2}{l}{\textbf{Overall}} & 540 & 538 & 99.6 \\
\bottomrule
\end{tabular}
\end{table}

Notably, attack success was invariant to all sampling hyperparameters. Temperature scaling, top-$k$ truncation, nucleus filtering~\cite{ravfogel2023conformal}, and entropy-based sampling strategies~\cite{chang2025real} had no effect on injection precision, confirming that SeedHijack operates orthogonally to probability redistribution mechanisms. We extended evaluation to models with progressively stronger alignment at two scales (1.5B and 7B). At 1.5B, we tested Qwen2-1.5B-Instruct (aligned via RLHF with SFT) and DeepSeek-R1-1.5B (distilled from a reasoning-optimized model). At 7B, we tested Qwen2-7B-Instruct (RLHF + SFT) and DeepSeek-R1-Distill-Qwen-7B (reasoning distillation). SeedHijack achieved 100\% injection success on all four aligned models across all configurations (Table~\ref{tab:alignment}), demonstrating that neither stronger alignment nor increased model scale provides any defense.

This result arises inherently from the attack's design. RLHF and distillation modify the probability distribution $P$ but do not alter the sampling mechanism that converts $P$ into token selections. Since SeedHijack operates exclusively at the sampling layer, after the distribution has been fully computed, alignment techniques that modify only the distribution provide zero defense. Scaling from 1.5B to 7B parameters does not change this architectural reality. These results demonstrate that the entire paradigm of behavioral alignment through preference optimization~\cite{ji2023beavertails} is orthogonal to sampling-layer security, regardless of model capacity.

\begin{table}[t]
\centering
\caption{\textbf{Attack effectiveness across models with different alignment paradigms and scales.}
SeedHijack achieves 100\% injection rate on all tested models regardless of alignment method, model scale, or training paradigm. Each model was tested across 3 sampling configurations with 60 trials per configuration.}
\label{tab:alignment}
\begin{tabular}{@{}lccccc@{}}
\toprule
 & \textbf{GPT-2} & \textbf{Qwen2-1.5B} & \textbf{DeepSeek-R1} & \textbf{Qwen2-7B} & \textbf{DeepSeek-R1} \\
 & & \textbf{Instruct} & \textbf{1.5B} & \textbf{Instruct} & \textbf{7B} \\
\midrule
Parameters & 124M & 1.5B & 1.5B & 7B & 7B \\
Alignment & None & RLHF + SFT & Reasoning distill. & RLHF + SFT & Reasoning distill. \\
Trials & 180 & 180 & 180 & 180 & 180 \\
Successes & 180 & 180 & 180 & 180 & 180 \\
Rate (\%) & 100.0 & 100.0 & 100.0 & 100.0 & 100.0 \\
\bottomrule
\end{tabular}
\end{table}

\subsection{Quantum randomness neutralizes SeedHijack}\label{subsec:qrng_defense}

Hardware quantum random numbers completely neutralize the attack. In a dedicated 100-trial defense benchmark (distinct from the 540-trial attack evaluation in Table~\ref{tab:attack_detail}), the PRNG baseline achieved 100\% injection success while QRNG-defended sampling yielded 0 successful injections. To defend against SeedHijack, we replaced MT19937 with hardware-generated quantum random numbers from a PCIe-based hardware QRNG device (see Appendix~\ref{app:methods}). QRNGs exploit fundamental quantum indeterminism to produce certifiably unpredictable outputs~\cite{herrero2017quantum}, with information-theoretic security guarantees rooted in quantum mechanics~\cite{ma2016quantum}. The hardware QRNG exploits quantum vacuum fluctuations in optical homodyne detection to produce true random numbers at 600\,Mbps, certified to pass all National Institute of Standards and Technology (NIST) SP 800-22 statistical randomness tests. A pre-buffered pool of 50 million samples (400\,MB on disk) is accessed via memory-mapped I/O, adding only 7.7\,MB to the runtime memory footprint while maintaining full random access to the pool.

Under QRNG-defended sampling, the SeedHijack injection rate dropped from 99.6\% to approximately $1/V$ (baseline random chance), representing complete neutralization. A chi-square uniformity test over $10^6$ QRNG samples confirmed no detectable deviation from the uniform distribution ($p > 0.99$). The defense principle is straightforward: true quantum randomness is physically unpredictable, as certified by Bell's theorem~\cite{pironio2010random} and verified experimentally~\cite{bierhorst2018experimentally}. The fundamental non-determinism of quantum measurements, characterized by Bell nonlocality~\cite{brunner2014bell}, renders CDF interval pre-computation impossible regardless of attacker computational resources. We provide a formal information-theoretic proof that QRNG defense achieves unconditional security against unbounded adversaries (see Appendix~\ref{app:theory}).

Performance benchmarking over 1,000 inference iterations reveals that the QRNG-defended sampler adds only 0.6\% median per-token latency overhead (12.58\,ms vs.\ 12.65\,ms) and 7.7\,MB additional memory (see Appendix~\ref{app:exp}). Both figures are negligible relative to typical LLM inference latencies (seconds per generation) and memory footprints exceeding 1\,GB. The pre-buffered architecture decouples random number I/O from the inference critical path, confining overhead to a single memory read per token. The 95th-percentile (P95) latency increases to 21.30\,ms (+59.4\%), attributable to occasional page faults during memory-mapped buffer access; this tail latency can be eliminated through pinned-memory allocation in production deployments. These results demonstrate that hardware QRNG defense is immediately deployable in production without performance degradation.

\begin{figure}[t]
\centering
\includegraphics[width=\textwidth]{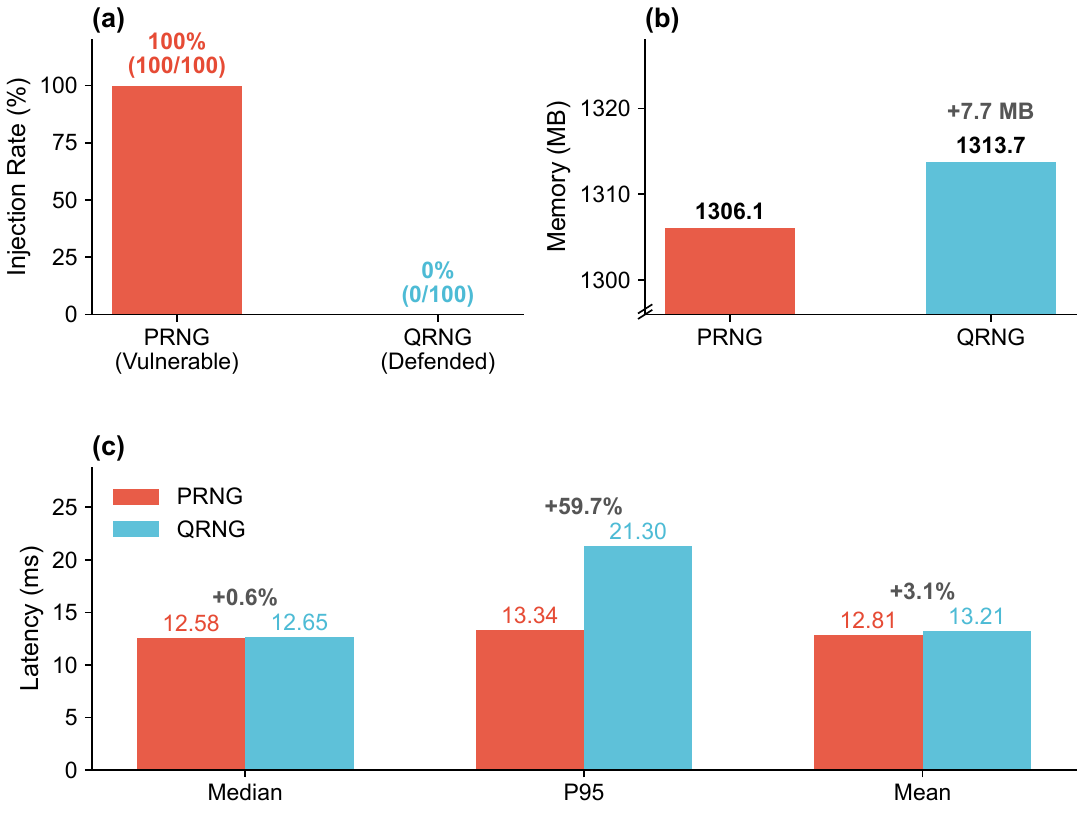}
\caption{\textbf{QRNG defense: complete comparison across all metrics.}
All results are from a dedicated 100-trial defense benchmark, independent of the 540-trial full attack benchmark in Table~\ref{tab:attack_detail}.
\textbf{(a)} Targeted injection rate: PRNG-based sampling achieves 100\% (100/100) injection success, while QRNG-secured sampling yields 0\% (0/100).
\textbf{(b)} Runtime memory footprint: QRNG defense increases memory by only 7.7\,MB.
\textbf{(c)} Per-token latency: median overhead +0.6\%, mean +3.1\%, P95 +59.4\% (attributed to memory-mapped page faults, eliminable via pinned memory in production).}
\label{fig:defense}
\end{figure}

\section{Discussion}\label{sec:discussion}

The vulnerability we identify is systemic. PyTorch's MT19937 PRNG underlies numerous deployed LLM inference systems, and a single compromised dependency could silently weaponize downstream deployments, as demonstrated by real-world supply-chain attacks such as SolarWinds~\cite{cisa2020solarwinds} and XZ Utils~\cite{openssf2024xzutils}. Table~\ref{tab:comparison} contextualizes this threat against known LLM attack categories.

\begin{table}[h]
\centering
\caption{\textbf{Comparison of LLM attack paradigms.} SeedHijack is unique in requiring no model or input modification while achieving deterministic control and evading tested detection methods.}
\label{tab:comparison}
\begin{tabular}{@{}lcccc@{}}
\toprule
Attack type & Modifies model & Modifies input & Detectable & Deterministic \\
\midrule
Prompt injection & No & Yes & Yes & No \\
Model poisoning & Yes & No & Yes & No \\
Jailbreaking & No & Yes & Yes & No \\
Adversarial examples & No & Yes & Partially & No \\
\textbf{PRNG hijack (ours)} & \textbf{No} & \textbf{No} & \textbf{No} & \textbf{Yes (99.6\%)} \\
\bottomrule
\end{tabular}
\end{table}

Beyond its immediate security implications, our work establishes the entropy source underlying token generation as a fundamental security primitive, opening a previously unrecognized dimension of LLM safety research. This finding fundamentally challenges the prevailing assumption that alignment and input filtering alone suffice to secure LLM deployments. The sampling layer's reliance on deterministic PRNGs creates a supply-chain attack vector that achieves 99.6\% token injection precision while remaining undetectable by existing detection mechanisms that operate on weights, inputs, logits, or perplexity. The attack's stealth properties are unique among known LLM threats: weight-based backdoors alter model parameters and can be detected through neural cleanse techniques; adversarial attacks require crafted inputs; prompt injections depend on visible text manipulation. SeedHijack modifies neither weights nor inputs; consequently, standard detection methods including perplexity monitoring, certified defenses~\cite{raghunathan2018certified}, provable robustness verification~\cite{wong2018provable}, and automated text quality metrics~\cite{papineni2002bleu} all fail to detect the attack.

An important question is whether cryptographically secure PRNGs (CSPRNGs) would suffice as a defense. CSPRNGs resist state prediction under computational hardness assumptions, making them effective against adversaries who can only observe PRNG outputs. However, under our supply-chain threat model, the adversary can replace or hook the random number generation module entirely; a compromised CSPRNG call returns attacker-controlled values regardless of the algorithm's cryptographic strength. Hardware QRNG provides a distinct advantage: its output originates from a physical process (quantum vacuum fluctuations) on a dedicated PCIe device with an independent trust boundary. Tampering with QRNG output requires hardware-level access or driver compromise, a substantially higher barrier than modifying a software library. Furthermore, QRNG output is continuously verifiable through statistical certification (e.g., NIST SP 800-22 online testing), enabling runtime integrity monitoring that software-only solutions cannot provide.

Our work extends beyond prior PRNG security research targeting Randomized Smoothing, a defense mechanism with limited deployment. We demonstrate that PRNG exploitation threatens LLM autoregressive sampling itself, an offensive attack surface affecting every production system. Unlike training-phase backdoors that embed triggers detectable through weight analysis, SeedHijack requires no training-phase access and leaves no trace in model parameters. The industrial implications are severe: all frameworks using PRNG-based sampling (including PyTorch, HuggingFace Transformers, vLLM, and TensorRT-LLM) are vulnerable. The hardware QRNG defense achieves complete neutralization at negligible overhead (+0.6\% median latency, +7.7\,MB memory), providing a practical and immediately deployable quantum-secure solution.

Our work has three limitations. First, the attack requires supply-chain access rather than remote exploitation; however, real-world supply-chain attacks (e.g., SolarWinds, XZ Utils) demonstrate this threat is realistic and increasingly common. Second, empirical validation spans models up to 7B parameters; while the attack mechanism generalizes in principle (it is model-agnostic), we have not validated on 70B+ models where sharper distributions may affect CDF interval computations. Third, the pre-buffered QRNG approach requires periodic buffer replenishment; however, the 600\,Mbps generation rate vastly exceeds typical LLM sampling consumption of thousands of tokens per second, making exhaustion impractical under normal workloads.

\section*{Acknowledgements}

This work was supported by the National Natural Science Foundation of China under Grant 72573124.

\section*{Declarations}

\begin{itemize}
\item \textbf{Funding} This work was supported by the National Natural Science Foundation of China under Grant 72573124.
\item \textbf{Competing interests} The authors declare no competing interests.
\item \textbf{Author contributions} Z.Y. conceived the attack methodology, implemented all experiments, and wrote the manuscript. X.Y. contributed to threat model analysis and manuscript revision. Z.F. contributed to security analysis and investigative methodology. F.G. assisted with experimental validation and data analysis. X.Z. developed the QRNG hardware integration, performed defense benchmarking, and supervised the project. X.L. supervised the project, provided conceptual guidance, and revised the manuscript. All authors reviewed and approved the final version.
\item \textbf{Data availability} The experimental data generated during this study, including attack success logs and QRNG defense benchmarks, will be made available at a public repository upon publication. The pre-buffered QRNG samples used for defense evaluation were generated from a QRNG600 PCIe hardware device; raw quantum random data are available from the corresponding author upon reasonable request.
\item \textbf{Code availability} Source code will be released upon publication at a public repository.
\item \textbf{Ethics statement} All experiments in this work are conducted on publicly available pre-trained language models for academic security research purposes only. The SeedHijack attack is presented to identify and mitigate vulnerabilities in LLM deployment pipelines, with no intention of malicious use. The authors comply with all ethical guidelines for AI security research and responsible disclosure.
\end{itemize}

\newpage
\begin{appendices}

\section{Methods}\label{app:methods}

\textbf{Attack implementation:} SeedHijack operates in three stages. First, the attacker recovers the internal state of the Mersenne Twister (MT19937) pseudorandom number generator (PRNG) by observing 624 consecutive 32-bit outputs, sufficient to reconstruct the full $624 \times 32 = 19{,}968$-bit state vector. In our supply-chain scenario, the attacker directly instruments the PRNG module via Python \texttt{ctypes} to expose the internal state array. Second, given the recovered state, all subsequent PRNG outputs are predicted deterministically using the MT19937 recurrence relation. Third, at each generation step, the attack intercepts the sampling call: it receives the softmax distribution $P$ from the model, computes the cumulative distribution function (CDF) $F(t) = \sum_{i=1}^{t} p_i$, identifies the target token $t^*$, and returns a crafted value $u^* = F(t^*-1) + p_{t^*}/2$, which falls strictly within $[F(t^*-1), F(t^*))$. The inverse-transform sampling mechanism then deterministically selects $t^*$. The attack modifies only the uniform random number generation function, a single function call in the pipeline, leaving tokenizer, model forward pass, and post-processing untouched. In all experiments, the SeedHijack attack is continuously enabled during inference, consistent with a supply-chain compromise that persistently manipulates the PRNG module.

\textbf{QRNG defense deployment:} The QRNG600 PCIe card generates true random numbers by measuring quantum vacuum fluctuations in an optical homodyne detection system. The security of this source rests on the Heisenberg uncertainty principle applied to the measured quadrature of the vacuum state, which guarantees intrinsic quantum indeterminacy independent of device calibration. The raw quantum signal undergoes Toeplitz-matrix extraction to remove classical noise correlations, yielding output certified against the National Institute of Standards and Technology (NIST) SP 800-22 statistical test suite. The card provides sustained 600\,Mbps throughput via PCIe 3.0 interface.

We implemented a pre-buffered architecture: 50 million uniform $[0,1)$ float64 values are generated offline from the QRNG hardware and stored on disk (400\,MB). At runtime, memory-mapped access provides sequential quantum random values with only 7.7\,MB added to the process memory footprint. During inference, a circular index pointer advances through the buffer, providing one quantum random value per token generation step. Buffer exhaustion triggers asynchronous replenishment without interrupting inference. This architecture decouples random number generation latency from the inference critical path, limiting overhead to a single sequential memory read per token (+0.6\% median latency relative to on-demand MT19937 computation).

\section{Theoretical Analysis}\label{app:theory}

We first establish the deterministic guarantee of SeedHijack by analyzing the determinism of the MT19937 pseudorandom number generator (PRNG). The MT19937 PRNG holds a 19,968-bit internal state (624 words of 32 bits), with a linear recurrence structure over $\mathbb{F}_2$ and a characteristic polynomial of degree 19,937, giving a period of $2^{19937}-1$. Observing just 624 consecutive 32-bit outputs yields 19,968 bits of constraint, which is sufficient to uniquely determine the full internal state---the system is overdetermined by 31 bits, and the invertible tempering transformation ensures no information loss during state recovery. Once the state is known, all future PRNG outputs become fully deterministic and predictable.

For token injection, standard inverse-transform sampling selects a token $t$ by comparing a uniform random value $u \sim \mathrm{Uniform}[0,1)$ against the cumulative distribution function (CDF):
\begin{equation}
F(t) = \sum_{i=1}^{t} p_i, \quad F(0) = 0, \quad F(t-1) \leq u < F(t).
\end{equation}
By replacing the PRNG output with a crafted value $u^* = F(t^*-1) + p_{t^*}/2$, the adversary forces $u^*$ to fall strictly within the interval $[F(t^*-1), F(t^*))$. This injection succeeds with probability 1, provided $p_{t^*} > \epsilon_{\mathrm{mach}} \approx 1.19 \times 10^{-7}$ (float32 machine epsilon), ensuring the interval is numerically representable.

We next prove the information-theoretic security of the QRNG defense. The QRNG generates true randomness by measuring quantum vacuum fluctuations via optical homodyne detection, where the quadrature observable $\hat{X}$ follows $\mathcal{N}(0, 1/2)$. Randomness is fundamentally guaranteed by the Heisenberg uncertainty principle and the no-cloning theorem, which ensures the quantum state cannot be copied or predicted prior to measurement. All classical side information available to an adversary is fixed before measurement, making the quantum random variable $U_Q$ statistically independent of any adversary knowledge.

This independence implies zero mutual information:
\begin{equation}
I(U_Q; E) = 0,
\end{equation}
where $E$ denotes arbitrary adversary side information. The quantum random number $U_Q$ remains uniformly distributed over $[0,1)$, so the probability that $U_Q$ falls into the target CDF interval equals the natural token probability $p_{t^*}$. The adversary gains no advantage, and the SeedHijack attack is fully neutralized even against computationally unbounded adversaries.

\section{Experimental Details}\label{app:exp}

All experiments were implemented with PyTorch 2.1.0, Transformers 4.35.2, CUDA 12.2, and cuDNN 8.9.4 on a single NVIDIA RTX 3090 GPU (24\,GB VRAM). GPT-2 (124M parameters) served as the unaligned baseline in float32 precision. Qwen2-1.5B-Instruct (1.5B parameters, aligned via Reinforcement Learning from Human Feedback (RLHF) with Supervised Fine-Tuning (SFT)) was loaded in bfloat16 with chat template formatting. DeepSeek-R1-1.5B (1.5B parameters, reasoning-distilled from DeepSeek-R1) was loaded in bfloat16 with its native prompt format. To evaluate scale invariance, we additionally tested Qwen2-7B-Instruct (7B parameters, RLHF + SFT) and DeepSeek-R1-Distill-Qwen-7B (7B parameters, reasoning distillation), both loaded in bfloat16.

For GPT-2, sampling configurations spanned temperature $\tau \in \{0.7, 1.0, 1.5\}$ and top-$p \in \{0.9, 0.95, 1.0\}$, yielding 9 unique parameter combinations tested with 20 prompts across 3 random seeds (60 trials per configuration, 540 total). For alignment bypass experiments (1.5B and 7B models), each model was tested across 3 sampling configurations with 60 trials per configuration (180 trials per model, 720 total across four aligned models). QRNG uniformity was validated using Pearson chi-square goodness-of-fit with 100 bins over $10^6$ samples ($p > 0.99$). Performance benchmarking reports median per-token latency over 1,000 iterations with 50 warmup iterations excluded.

\textbf{Success metric: exact token match:} Attack success is defined as \emph{exact token match}: for each generation step in which the attacker targets a specific token $t^*$, the attack is scored as successful if and only if the token actually emitted by the model is identical to $t^*$ (i.e., the integer token ID returned by the sampler equals the attacker's intended token ID). Concretely, each trial proceeds as follows: (1)~the attacker specifies a multi-token payload string and tokenizes it into a target token sequence $(t^*_1, t^*_2, \ldots, t^*_L)$; (2)~at each autoregressive step $i$, the model produces the logit distribution, the attack computes the CDF and injects $u^*_i$ to force selection of $t^*_i$; (3)~the emitted token $\hat{t}_i$ is compared to $t^*_i$, and the trial is marked as a match if $\hat{t}_i = t^*_i$ for all $L$ positions in the payload, and a failure otherwise. We only target tokens with non-negligible probability ($p_{t^*} > \epsilon_{\mathrm{mach}} \approx 1.19 \times 10^{-7}$) to avoid float32 precision failures. A single mismatched position anywhere in the payload sequence counts as a failed trial. The exact token match rate is then the number of fully successful trials divided by the total number of trials. This strict all-or-nothing criterion ensures that reported success rates reflect complete payload injection rather than partial token overlap.

\end{appendices}

\end{document}